\documentclass[a4paper,11pt]{article}
\usepackage{pos}
\usepackage{slashed}
\usepackage{amsmath}
\newcommand{\be}{\begin{equation}}
\newcommand{\ee}{\end{equation}}

\title{Torsion-induced axions in string theory,\\ quantum gravity and the cosmological tensions}

\author*[a,b]{Nick E. Mavromatos}
\author[a]{Panagiotis Dorlis}
\author[a]{Sotirios-Neilos Vlachos}

\affiliation[a]{National Technical University of Athens, School of Mathematical and Physical Sciences, Physics Division,\\
9 Iroon Polytechniou, Zografou Campus 157 80, Athens, Greece}

\affiliation[b]{King's College London, Department of Physics, Theoretical Particle Physics and Cosmology Group, \\Strand, London WC2R 2LS, London, UK}

\emailAdd{mavroman@mail.ntua.gr}

\abstract{We discuss the r\^ole of torsion in string theory on inducing pseudoscalar degrees of freedom (axions), which in turn couple to (gravitational) Chern-Simons (CS) anomalous terms. Such interactions can induce inflation, of running vacuum type, not requiring external inflaton fields, through condensation of the anomalous terms as a consequence of  primordial chiral gravitational-wave (GW) tensor perturbations in a weak-quantum gravity setting. The presence of an UV cutoff for the GW quantum graviton modes opens up the system, leading to a dissipative behaviour realised via the presence of non trivial imaginary parts of the gravitational CS terms. The naive estimate of the life time of inflation based on such imaginary parts, which afflict the pertinent GW Hamiltonian, is quite consistent with the estimates of the duration of inflation based on an analysis of the condensate-induced linear-axion-potential by means of dynamical systems.  
Such quantum-gravity effects can also contribute positively to the alleviation of cosmological tensions if they survive today. In the talk we discuss the conditions under which such a result may be achieved. We also discuss the potential r\^ole of other axions in string theory, coming from compactification, in inducing enhanced densities of primordial black holes during RVM inflation, thereby contributing to significantly increased percentages of these black holes that can play the r\^ole of dark matter components. Moreover, under certain circumstances, that we shall discuss in some detail, it is also possible that the initially massless torsion-induced axions can acquire a non-trivial mass during the radiation era, thereby providing additional dark matter components in the Universe.  With regards to this aspect, we also emphasise the r\^ole of massive right-handed neutrinos, provided that such excitations exist in the relevant spectra.}

\FullConference{Corfu Summer Institute 2023 "School and Workshops on Elementary Particle Physics and Gravity" (CORFU2023)\\
 23 April - 6 May , and 27 August - 1 October, 2023\\
Corfu, Greece\\}

\begin{document}
\maketitle

\section{Introduction and summary}

Axion (pseudoscalar) fields play an important r\^ole in theoretical cosmological models~\cite{Marsh:2015xka}, although at present they have not been discovered yet. In string theory they arise naturally~\cite{Svrcek:2006yi}, either in the massless gravitational multiplet, as duals (in the (3+1)-dimensional spacetime after string compactification) of the field strength of the spin-1 antisymmetric tensor Kalb-Ramond (KR) field, or due to compactification. The former axions appear in all string theories, and they are independent of the details of compactification (this is why they are also called string-model independent axions). The latter obviously depend on the specific string theory and the associated compactified geometry from which they arise. 

In the present talk I will mostly deal with the string model independent (or KR) axions in effective (3+1)-dimensional cosmologies, obtained after generic string compactification. These cosmologies exhibit interesting features, such as:  inflation of running-vacuum-model~\cite{solarvm} (RVM) type~\cite{Basilakos:2019acj,Mavromatos:2020kzj}, without the need for external inflaton fields, CPT Violating (CPTV) leptogenesis during the early radiation era, of the type discussed in \cite{Mavromatos:2012ii,deCesare:2014dga,Bossingham:2017gtm,Bossingham:2018ivs,Mavromatos:2020dkh}, and also, 
the possibility, under some circumstances that we shall discuss here, of simultaneous alleviation of the Hubble $H_0$ and growth-of-structure tensions~\cite{Gomez-Valent:2023hov}, observed in the current era cosmological data, when the pertinent observations are confronted with theoretical predictions coming from $\Lambda$CDM cosmology~\cite{Planck:2018vyg}. The latter feature occurs under the inclusion of {\it quantum-gravity corrections} (in a weak gravity setting). Indeed, quantum-graviton one-loop corrections to the effective action, when one expands around de-Sitter epochs, such as the inflationary or modern eras in the Universe cosmic history, result in terms of the form $H^{2n} \ln(H)$, $n \in \mathbb Z^+$. In modern eras, the terms with $n=2$ dominate, and in fact are those which are held responsible for the simultaneous alleviation of both types of cosmological tensions~\cite{Gomez-Valent:2023hov}.

All the above features concern a specific type of interactions that characterises these string-inspired cosmological models, namely interactions of the KR axions with the topological CP-violating gravitational Chern-Simons (gCS) term (or Hirzenbruch signature~\cite{Duncan:1992vz}). The presence of such terms is a characteristic feature of strings~\cite{Green:1987sp, Green:1987mn}, due to modifications in the definition of the KR field strength as a consequence of the Green-Schwarz anomaly cancellation mechanism~\cite{Green:1984sg}. 
Essentially the model is a Chern-Simons (CS) gravity~\cite{Jackiw:2003pm}, which arises when one truncates to second order ({\it i.e} $\mathcal O(\alpha^\prime)$, where $\alpha^\prime$ the Regge slope) the infinite series in spacetime derivatives that characterises the low-energy effective field-theory limit of strings. The inflationary epoch arises from condensation of the gCS terms due to primordial gravitational waves (GW)~\cite{Mavromatos:2020kzj}, and is due to $H^4$ terms in the effective vacuum energy density, where $H$ denotes the (approximately) constant Hubble parameter during inflation. It is such non-linear $H^4$ terms  the lead to an early inflationary phase in the cosmological RVM evolution~\cite{Lima:2013dmf, Perico:2013mna} this model is subject to, which does not require external inflaton fields. 

An important remark we would like to make at this point concerns the role of the KR field strength in $\mathcal O(\alpha^\prime)$ string-effective actions as a (totally antisymmetric part of a) torsion~\cite{Hehl:1976kj,Shapiro:2001rz,Mavromatos:2021hai,Mavromatos:2023wkk}. In this respect, the CS gravity models also characterise generic Einstein-Cartan theories with fermions in contorted geometries, not related to strings, where the totally antisymmetric part of the torsion is associated with the derivative of a pseudoscalar (axion-like-particle (ALP)) degree of freedom~\cite{Duncan:1992vz}.

During inflation in the string-inspired RVM model of \cite{Basilakos:2019acj,Mavromatos:2020kzj},
called from now StRVM, one has a linear axion potential~\cite{Dorlis:2024yqw}, as a consequence of the condensate, whose presence also guarantees a ground state configuration of the KR axion field, linear in cosmic time, which violates spontaneously Lorentz and CPT symmetry. Under certain conditions, this axion configuration can survive undiluted the inflationary epoch,  and leads to the aforementioned CPTV leptogenesis in the radiation epoch in theories involving massive right-handed neutrinos (RHN)~\cite{Mavromatos:2012ii,deCesare:2014dga,Bossingham:2017gtm,Bossingham:2018ivs,Mavromatos:2020dkh}. The latter can then be communicated to the baryon sector via sphaleron processes in the standard model sector of the effective theory, which preserve the difference of the Baryon minus Lepton numbers. In this way, an unconventional mechanism for matter-antimatter asymmetry in the universe is obtained, which has its routes in the underlying geometry of strings, given that the KR axion is associated with torsion, as already mentioned.

The structure of the talk is as follows: in the next section \ref{sec:model}, we discuss the effective string-theory model of CS gravity with torsion-induced KR axions, and explain how a linear axion potential arises due to condensation of the anomaly terms, as a result of primordial GW-modes, which are generated in a pre-RVM inflationary epoch of this universe. We argue, using a dynamical-system approach, that such potentials lead to metastable inflationary epochs, 
and estimate the duration of inflation, which depends on the initial conditions of the cosmological axion system. As an independent check, we perform an alternative estimate of the inflation-era life time by making use of the imaginary parts of the CS condensate, and, hence, of the Hamiltonian describing the gravitational dynamics induced by the chiral GW, and find consistency with the dynamical-system estimate which was based on the real part of the CS condensate. We note that the presence of imaginary parts in the computation of the gravitational CS condensate is due to the imposition of an UV cutoff in the momenta of the pertinent quantum graviton modes, associated with the (weak) GW pereturbations. Such a cutoff, opens up the quantum-graviton system, with the modes with momenta above the cutoff playing the r\^ole of an ''environment'', which induces dissipation, realised through the presence of the aforementioned imaginary parts. In the current work we make use of such instabilities to argue on the metastable nature of the RVM inflationary era in the StRVM. We note at this stage that  a rather different instability of the CS gravity, upon the imposition of an UV cut - off on the GW spectrum, has been observed in \cite{Dyda:2012rj} in a rather different perspective, by including ghost graviton modes. Par contrast, in our case, as explained in \cite{Dorlis:2024yqw}, and mentioned later on in this talk, such ghost gravitons are absent from our spectrum, due to the choice of the UV cutoff.

We also discuss the periodic modulation of the axion potentials in multiaxion situations encountered in strings, which involve axions arising from compactification in addition to the KR axions. Non-perturbative world-sheet instantons are held responsible for the generation of such periodic modulations of the axion potentials, which in turn may affect the densities of the primordial black holes during inflation, and, as a consequence, the profile of GW in the early radiation era. In section \ref{sec:tensions}, we discuss the features of the model in post-RVM inflationary eras, in particular 
the generation of KR axion potentials, including mass terms, during the early radiation epoch, either through the path-integration of heavy RHN, or through quantum-chromodynamics (QCD) instanton effects. Both effects are the result of the breaking of the shift symmetry, either by the mass of the RHN or by the QCD instantons. In this way, the so arising massive KR (torsion-induced) axions could play the role of a dark matter component, which is thus geometrical in origin. In modern epochs, including the present era, we discuss one-loop quantum-graviton effects around the spacetime background of the modern de-Sitter era of the Universe, and argue that, under some circumstances, these effects can dominate the ones encountered in matter quantum field theories in such curved spacetimes, and be responsible for the alleviation of  both types of cosmological tensions, provided the latter have a fundamental physics origin, and are not artefacts of statistics or admit mundane astrophysical explanations~\cite{Freedman:2017yms}. Finally, section \ref{sec:concl} contains our conclusions and a brief outlook.

\section{The Stringy Running Vacuum Model and Inflation: the r\^ole of torsion-induced axions}\label{sec:model}

Our starting point is the effective (3+1)-dimensional string-inspired gravitational theory with axions and CS anomalies, derived in refs.~\cite{Basilakos:2019acj,Mavromatos:2020kzj}, where we refer the interested reader for details.
The effective cosmological field-theory model is based on the (bosonic sector of the) massless ground state  of the underlying superstring theory, and to $\mathcal O(\alpha^\prime)$ is given by:\footnote{Throughout this work we  work in units $\hbar=c=1$, and use the following  metric-signature conventions and definitions: signature of metric $(-, +,+,+ )$, Riemann Curvature tensor 
$R^\lambda_{\,\,\,\,\mu \nu \sigma} = \partial_\nu \, \Gamma^\lambda_{\,\,\mu\sigma} + \Gamma^\rho_{\,\, \mu\sigma} \, \Gamma^\lambda_{\,\, \rho\nu} - (\nu \leftrightarrow \sigma)$, Ricci tensor $R_{\mu\nu} = R^\lambda_{\,\,\,\,\mu \lambda \nu}$, and Ricci scalar $R_{\mu\nu}g^{\mu\nu}$. } 
\begin{align}\label{sea3}
S^{\rm eff (I)}_{\rm B} =&\; \int d^{4}x\sqrt{-g}\Big[ \dfrac{1}{2\kappa^{2}}\, R - \frac{1}{2}\, \partial_\mu b \, \partial^\mu b - \sqrt{\frac{2}{3}} \, \frac{\alpha^\prime}{96\, \kappa} \, b(x) \, R_{\mu\nu\rho\sigma}\, \widetilde R^{\nu\mu\rho\sigma}  + \dots \Big],
\end{align}
where $\kappa^2= M_{\rm Pl}^{-1}$ is the (3+1)-dimensional gravitational coupling, with $M_{\rm Pl} \simeq 2.4 \cdot 10^{18}~{\rm GeV}$ the reduced Planck mass,  $\alpha^\prime =M_s^{-2}$ is the Regge slope, with $M_s$ the string mass scale, which in general is different from $M_{\rm Pl}$, and is treated as a phenomenological parameter, and the dots $\dots$ denote higher-derivative  and gauge terms, which are not relevant for our purposes in this section.  In \eqref{sea3}, the symbol $\widetilde{R}_{\alpha\beta\gamma\delta}=\frac{1}{2}R_{\alpha\beta}^{\,\,\,\,\,\,\,\,\rho\sigma}\varepsilon_{\rho\sigma\gamma\delta}$, denotes the dual of the Riemann tensor in (3+1)-dimensional spacetimes, with 
$\varepsilon_{\rho\sigma\kappa\lambda} = \sqrt{-g(x)} \, \hat \epsilon_{\rho\sigma\kappa\lambda} $  the covariant Levi-Civita symbol, under the convention that the symbol $\hat \epsilon_{0123}=1$, {\it etc} is the  totally antisymmetric symbol  in (3+1)-dimensional flat-Minkowski spacetime. The field $b(x)$ is the KR, or string-model independent, axion-like-particle~\cite{Duncan:1992vz,Svrcek:2006yi}, whose derivative is associated with the dual (in (3+1)-dimensional spacetimes) of the field strength of the spin-one antisymmetric tensor of the string ground-state multiplet. In \eqref{sea3} we have set the dilaton to a constant value, which can be self-consistently guaranteed in the context of low-energy field theories obtained from strings, upon inclusion of an appropriate dilaton potential, which is minimised accordingly~\cite{Basilakos:2020qmu}. Such a dilaton potential can arise from string-loop corrections to the effective action.

In~\cite{Basilakos:2019acj,Mavromatos:2020kzj}, it was proposed that primordial GW, which can be generated in a pre-RVM inflationary phase of the model \eqref{sea3} characterised by stiff-KR-axion dominance~\cite{Mavromatos:2020kzj,Mavromatos:2021urx}, can condense leading to a condensate of the CS anomaly term: 
\begin{align}\label{CScond}
\langle R_{\mu\nu\rho\sigma}\, \widetilde R^{\nu\mu\rho\sigma} \rangle_{{\mathcal N}_I} \simeq {\rm constant}\,,
\end{align}
where $\simeq$ denotes an approximate constancy, that is, at most a very mild variation with the cosmic time is allowed,
in the isotropic and homogeneous expanding universe background of approximately de Sitter type which characterises the 
RVM-inflationary epoch of the model. In \eqref{CScond}
the index ${\mathcal N}_I$ 
denotes the number density per proper volume of primordial GW sources. Under the assumption of a linear superposition of the GW perturbations due to the various sources, one obtains~\cite{Mavromatos:2022xdo} that the condensate \eqref{CScond} is proportional to ${\mathcal N}_I$. 

The existence of a non-zero condensate in the presence of CS terms is guaranteed if chiral (that is left, right mode asymmetric~\cite{Lue:1998mq}) GW (quantum) perturbations of the background metric are taken into account,\footnote{For scenarios of such primordial GW generation we refer the reader to \cite{Mavromatos:2020kzj}, and references therein.} in a weak-quantum-gravity setting~\cite{Alexander:2004us,Lyth:2005jf}, where graviton modes are integrated out up to a momentum-scale Ultraviolet (UV) cutoff $\mu$. In the context of low-energy effective field theories obtained from strings the cutoff $\mu$ can be naturally identified with the string scale $M_s$~\cite{Mavromatos:2022xdo}. The latter is a phenomenological parameter, which can be taken to be at most equal to the reduced Planck scale, $M_s \le M_{\rm Pl}$, so that the transplanckian censorship hypothesis is satisfied, which in this context means that no graviton modes are allowed to have momenta with a magnitude exceeding the Planck scale. Then, it can be shown~\cite{Basilakos:2019acj,Mavromatos:2020kzj} that the GW-induced condensates  \eqref{CScond} do survive the exponential expansion during the RVM inflation, provided a sufficiently large number of GW sources exist during the RVM inflationary epoch~\cite{Mavromatos:2022xdo}. The abrupt formation of these sources, in short time periods relative to the duration of inflation, is responsible for the transition from the stiff - axion - era to the RVM inflationary phase~\cite{Dorlis:2024yqw}. The so-required number of such sources is expressed as a function of the ratio $M_s/M_{\rm Pl}$.
We stress once again that this approach is compatible with setting the UV cutoff of the GW modes equal to the string scale $M_s$, and keeping $M_s \lesssim M_{\rm Pl}$, as a phenomenological parameter.

The computation of the condensate proceeds via weak-quantum-gravity perturbations, of chiral GW type, of the Friedmann-Lemaitre-Robertson-Walker (FLRW) background spacetimes. The action describing the graviton modes produced though the chiral GW reads~\cite{Dorlis:2024yqw}, 
\begin{equation}
    S=-\frac{1}{2\kappa^2}\sum_{\lambda=L,R} \int d\eta\int d^3k \;z_{\lambda,\vec{k}}^2(\eta)\left( \vert\widetilde{h}_{\lambda,\vec{k}}^\prime\vert^2 -k^2 \vert\widetilde{h}_{\lambda,\vec{k}}\vert^2\right)\,,
    \label{GWAction}
\end{equation}
where 
\begin{equation}
    z_{\lambda,\vec{k}}(\eta)=\alpha \sqrt{1 -l_{\lambda} l_{\vec{k}}  L_{CS}(\eta)}\,,\;\; l_\lambda=\pm1,\;\lambda=L,R
    \label{zdef}
\end{equation}
and 
\begin{equation}
    L_{CS}(\eta)=k  \xi\ , \qquad     \xi=  \frac{4 A  b^{\prime}\kappa^2}{\alpha^2}\,, \qquad A = \sqrt{\frac{2}{3}} \, \frac{\alpha^\prime}{48\, \kappa}\,,
    \label{L_CS}
\end{equation}
and the prime indicates derivative with respect to the conformal time $\eta$ of the FLRW background spacertime.
Thus, the condition for the redefinition function, $z_{\lambda,\vec{k}}$, to be real, so that {\it ghost instabilities} are {\it avoided}, sets a lower limit for the UV cut-off, $\mu = M_s$, at the order of:\footnote{For the calculation of the bound \eqref{lbound}~the approximately constant value for $\dot b$:
\begin{align}\label{bdot}
\dot{b}\approx 10^{-1}H_IM_{\rm Pl}\,,
\end{align}
during the RVM inflation, 
has been taken into account~\cite{Dorlis:2024yqw}, where the inflationary-era Hubble parameter $H_I$ is set to the value:
\begin{align}\label{HInfl} 
H_I\approx10^{-5}M_{\rm Pl}\,,
\end{align}
as follows from the current data fits of single-field inflationary models~\cite{Planck:2018vyg}. Such a value for $\dot b$ is compatible with an inflationary epoch that lasts for 50-60 e-folding, as follows from the dynamical system analysis of the linear-axion potential performed in \cite{Dorlis:2024yqw}. The value \eqref{bdot} is also in agreement with the analysis of \cite{Basilakos:2019acj,Mavromatos:2020kzj}.}
\begin{align}\label{lbound}
M_s\;>\;10^{-8}M_{\rm Pl}\,, 
\end{align}
compatible with the aforementioned transplancian conjecture.

The presence of the condensate \eqref{CScond} leads to a {\it linear} KR axion potential, of axion-monodromy type, that is encountered in some string/brane-inspired cosmological models, after appropriate compactification~\cite{McAllister:2008hb}. Such a linear-axion potential can lead to inflation. However in such a conventional string/brane theory context, the coefficient of the linear axion field is a constant depending on the parameters of the appropriate string compactification used. Par contrast, in the StRVM the condensate has structure, and in fact it contains terms scaling with the quadratic and quartic powers of the Hubble parameter  $H(t)$, in the appropriate epoch in which it is computed (where $t$ denotes the cosmic time of a Robertson-Walker observer). 
In fact, within our context, by calculating the CS condensate \eqref{CScond} at the end of the stiff-axion era, which precedes the RVM inflation~\cite{Mavromatos:2020kzj}, 
one can use the dynamical-system approach~\cite{Dorlis:2024yqw} to study the details of the passage from that era to the RVM inflation, utilising the (approximate) linearity of the axion potential. However, in the StRVM, whose dynamics is expressed by the model \eqref{sea3}, the dominant driving force for generating an inflationary era of RVM type~\cite{Perico:2013mna,Lima:2013dmf} is the quartic power of $H^4$ that appears in the effective action, and thus in the vacuum energy density, due to the CS condensate~\cite{Basilakos:2019acj,Mavromatos:2020kzj}. In fact, it can be explicitly shown~\cite{Mavromatos:2020kzj} that the equation of state (EoS) in the phase of a CS-condensate dominance in the model \eqref{sea3} is $w=-1$, despite the mild cosmic-time $t$ dependence of the vacuum energy density on the cosmic time through $H(t)$. This is exactly the situation encountered in the RVM framework~\cite{solarvm}, and does not characterise the linear-axion-induced inflation in conventional strings. Nonetheless, as demonstrated by the explicit computations of \cite{Dorlis:2024yqw}, the linear axion does induce the appropriate transition from a stiff-axion to an inflationary phase, but the latter is dominated by the non-linear $H^4$ terms appearing in the CS condensate. 

In the case of the linear axion potential induced by the condensate of chiral GW, inflation arises as a saddle point of the cosmic evolution~\cite{Dorlis:2024yqw}. It should be stressed at this point that the passage from such a point is achieved only under certain initial conditions for the axion field, corresponding to the time that the condensate is formed during the stiff - axion - era that preceeds the RVM inflationary era~\cite{Mavromatos:2020kzj}. If such initial conditions occur, the passage of the system from the saddle point induces an appropriate nearly-de-Sitter expansion, with an effective cosmological constant given by,     
\begin{equation}
    \Lambda_{eff}= \frac{1}{2}A \langle R_{\mu\nu\rho\sigma}\widetilde{R}^{\nu\mu\rho\sigma}\rangle_{\mathcal{N_I}} b\approx {\rm const} \, > \, 0 ~,
\label{effectivecosmologicalconstant}
\end{equation}
with $A$ defined in \eqref{L_CS}. 
The reader should notice that the behaviour of the axion \eqref{bdot} is crucial for ensuring the approximate constancy (in order of magnitude) of $\Lambda_{eff}$ during the entire duration of the RVM inflation $\Delta t_I$, provided at the beginning of inflation the axion $|b(t_{\rm begin})| \sim 10\, M_{\rm Pl}$, which is compatible with the initial conditions for inflation in the analysis of \cite{Dorlis:2024yqw}, but also of \cite{Basilakos:2019acj,Mavromatos:2020kzj}.
It is also important to notice that the time of formation of the condensate during the stiff - axion - era is essential~\cite{Dorlis:2024yqw} for obtaining the phenomenologically viable duration of inflation, $H_I \Delta t_I \sim N_e = \mathcal O(50-60) $, where $N_e$ is the number of e-foldings, with the Hubble parameter during inflation $H_I$ assumed to take on the phenomenological value \eqref{HInfl}, as inferred from the data~\cite{Planck:2018vyg}. Such a dependence on the initial conditions is an artifact of the {\it spontaneous breaking} of the shift symmetry for the axion $b(x)$, due to the formation of the condensate, while the "saddle-point'' nature of inflation induces the appropriate exit from the inflationary era, making the RVM de-Sitter era a {\it metastable phase} of the evolution of the StRVM. This is crucial for ensuring the consistency of the StRVM with the swampland criteria~\cite{swamp1,swamp2,swamp3,swamp4,swamp5}.

 It is important to mention at this point that the metastability of the de Sitter vacuum characterising the RVM inflation is also compatible with 
 the existence of appropriate imaginary parts in the effective action, arising in the weak quantum-gravity treatment of \cite{Lyth:2005jf,Dorlis:2024yqw}. Indeed, the gravitational CS-condensate, beyond its real part that contributes to the effective cosmological constant \eqref{effectivecosmologicalconstant} via the appropriate contribution of the sources, acquires additionally an imaginary part, given by:\footnote{Note that here $\langle \dots \rangle$ denotes the vacuum expectation values for a single source of GW, unlike the case of the real part of the CS condensate \eqref{effectivecosmologicalconstant}, where all sources contributions are  additive. The instability of the GW  vacuum, associated with the existence of a non-trivial imaginary part \eqref{ImCS}, follows by considering a single source of GW.} 
\begin{equation}\label{ImCS}
    {\rm Im} \left(\langle R_{\mu\nu\rho\sigma}\widetilde{R}^{\nu\mu\rho\sigma}\rangle \right)= \frac{16  A \text{$\dot{b}$} \mu ^7}{7 M^{4}_{\rm Pl} \ \pi ^2} \left[1+\left(\frac{H_I}{\mu}\right)^2\left(\frac{21}{10}-6\left(\frac{A\mu\text{$\dot{b}$}}{M_{\rm Pl}^2}\right)^2\right)\right]
\end{equation}
where the last term 
on the right-hand side of the above equation is subdominant, but we give it here explicitly, for completeness. The imaginary parts of the condensate back-react also on the effective Lagrangian and, hence, the corresponding Hamiltonian $\mathcal H$ also acquires an imaginary part, which can be estimated as follows:
\begin{equation}
 {\rm Im}\left(\mathcal H\right) = \int d^3 x \ \frac{1}{2} A \ b \ {\rm Im} \left(\langle R_{\mu\nu\rho\sigma}\widetilde{R}^{\nu\mu\rho\sigma}\rangle \right) \approx V^{(3)}_{dS} \ \frac{8  b A^2  \text{$\dot{b}$}  \mu ^7}{7 M^{4}_{\rm Pl} \ \pi ^2}~.
\label{imaginaryhamiltonian}
\end{equation}
where $V^{(3)}_{dS}$ denotes the de Sitter 3-volume. 

We stress at this point that the imaginary parts of the CS condensate arise due to the presence of the UV cutoff $\mu$, which in turn implies that the system of graviton modes with momenta below $\mu$ is an open, dissipative quantum system interacting with an ``environment'' of the modes with momenta above the cutoff.\footnote{In this sense, the dissipative 
nature of our graviton system is somewhat analogous to the representation of a decaying particle system as an open quantum system, where the states of the decay products play the r\^ole of the pertinent ``environment'', see, for instance the works in \cite{Caban:2005ue,Bertlmann:2006fn,Bernabeu:2012au}.} The dissipative nature of our open-system, due to the introduction of a cut-off $\mu$ in our effective description, provides us with a naively defined estimate of the lifetime $\tau$ of the inflation vacuum (in natural units, where $\hbar=1$):\footnote{The estimate \eqref{lifetime} is considered naive because of the fact that the inflationary vacuum in the StRVM is a complicated quantum-gravitational ground state, and thus the naive estimate of its lifetime, as given by the inverse of the imaginary part of the respective Hamiltonian, which is the standard lore in the case of unitary quantum mechanical or particle systems, may not be strictly valid. Nonetheless, as we shall see below, such a naive estimate leads to consistent results on the duration of inflation with those of the dynamical-system approach of \cite{Dorlis:2024yqw}.}
\begin{align}\label{lifetime}
\tau \sim \left(
{\rm Im} \mathcal{H} \right)^{-1}\,.
\end{align}
Note that the (Euclidean) four volume $V^{(3)}_{dS} T^E$ of the de-Siter spacetime, is given by \cite{Fradkin:1983mq, Mavromatos_decay1,Mavromatos_decay2}
\begin{equation}
   V^{(3)}_{dS} T^E =  \frac{24 \pi^2}{M^{2}_{\rm Pl} \Lambda} \ , \quad \Lambda \approx 3 H^{2}_I~,
   \label{4volume}
\end{equation}
with $T^E$ corresponding to the Euclidean time, defining the appropriate duration of inflation via 
\begin{align}\label{durinfl}
T^E\sim (50-60)H_I^{-1}\,. 
\end{align}
Substituting the above to eq.\eqref{imaginaryhamiltonian}, we obtain:
\begin{equation}\label{ImHb}
    {\rm Im}\left(\mathcal H\right) = \frac{64 b A^2 \text{$\dot{b}$} \kappa ^6 \mu ^7}{ 7 H_I }\cdot\frac{1}{H_I T^E}~.
\end{equation}
The order of magnitude of the quantities entering \eqref{ImHb} can be taken from \cite{Dorlis:2024yqw}, by leaving the string-scale $M_s = \mu$ as a free parameter. Then, we can easily obtain that the (naively defined) lifetime $\tau$ of the RVM inflationary vacuum of the StRVM is given by: 
\begin{equation}
    H_I\tau\sim \frac{7H_I^2M_{\rm Pl}^6}{64bA^2\dot{b}M_s^7}\left(H_I T^E\right) \sim 10^{-2}\left(\frac{M_{\rm Pl}}{M_s}\right)^{3} \cdot (H_I T^E)~.
\end{equation}
For such a decay rate to be consistent with the duration of inflation \eqref{durinfl}, the following upper limit for the string-effective-theory UV cut-off $\mu=M_s$ has to be satisfied:
\begin{equation}
\frac{M_s}{M_{\rm Pl}}\lesssim
0.215~~.
\end{equation}
This is quite consistent with the findings of the linear-axion-potential dynamical-system analysis of \cite{Dorlis:2024yqw}, which concentrated on the real part of the CS condensate. 
The above discussion leads to the conclusion that the r\^ole of KR axions (or other string-compactification-induced axions, which also couple to the gravitational CS term) in inducing RVM inflation is instrumental, since, in their absence, the CS term, being a total derivative, would not contribute to the dynamics in \eqref{sea3} and the pertinent condensate would be trivially vanishing.

In addition to their r\^ole in inducing RVM inflation, string axions play another important r\^ole. Non-perturbative world-sheet instanton effects may generate periodic modulation on the axion potentials~\cite{McAllister:2008hb}, in addition to the linear axion contributions coming from the CS condensates. In multiaxion-models, such as the ones expected to correspond to realistic string theories, in which both KR and string-compactification-induced axions are present, depending on the relevant parameters, one may face a situation in which the periodic modulations due to the compactification axions dominate the late stage of the KR-axion-induced RVM-inflation, thus producing enhanced densities of primordial black holes at the inflation-exit phase, which in turn will modify the profiles of the GW generated, e.g., by the non-spherically symmetric merging of these PBHs, and thus the GW spectrum in the early radiation era~\cite{Mavromatos:2022yql}. In fact, such effects can be in principle distinguishable, e.g. in future interferometers such as LISA, from conventional string theory models of axion monodromy, such as those in~\cite{Zhou:2020kkf}. 

Specifically, assuming for simplicity and concreteness that only one compactification axion is dominant during the RVM inflation, in addition to the torsion-induced KR axion,  the situation is effectively described by the following effective potential
in the axionic sector~\cite{Mavromatos:2022yql}:
\begin{align}\label{effpot}
 V(a, \, b) &={\Lambda_1}^4\left( 1+ f_a^{-1}\, \tilde \xi_1 \, a(x) \right)\, \cos({f_a}^{-1} a(x))+\frac{1}{f_{a}}\Big(f_b \, {\Lambda_0}^3 + \Lambda_2^4 \Big) \, a(x) + {\Lambda_0}^3\, b(x), \nonumber \\
 & \Lambda_0 \propto \langle R_{\mu\nu\rho\sigma}\, \widetilde R^{\nu\mu\rho\sigma} \rangle\,,
 \end{align}
 where $f_a$ is the compactification-axion-$a$-field coupling; the KR axion coupling $f_b$ is given by the inverse of the 
 numerical coefficient in front of the gCS term (third term on the right-hand-side) of the action \eqref{sea3}, divided by the 
 normalisation factor $16\pi^2$, that is 
 \begin{align}\label{kraxcoupl}
f_b = \frac{96}{16\, \pi^2} \, \sqrt{\frac{3}{2}} \, \frac{\kappa}{\alpha^\prime} \simeq 1.3 \, \frac{\kappa}{\alpha^\prime} = 
1.3 \, \frac{M_s^2}{M_{\rm Pl}}\,.
\end{align}
This can be understood as follows: the gCS term is only part of the structure arising from the implementation of a Bianchi identity of the modified field strength of the antisymmetric tensor field due to the Green-Schwarz mechanism~\cite{GS}. In the presence of (non-Abelian in general) gauge fields, the complete term in the effective action, replacing the gCS term would assume the mixed (gauge and gravitational) anomaly form 
\begin{align}\label{mixed}
S^{\rm eff (\rm anomalies)}_{\rm B} \ni &\; - \int d^{4}x\sqrt{-g}\, \sqrt{\frac{2}{3}} \, \frac{\alpha^\prime}{96\, \kappa} \, b(x) \, \Big[ R_{\mu\nu\rho\sigma}\, \widetilde R^{\nu\mu\rho\sigma}  + \rm tr \Big( \mathbf F^{\mu\nu}\, \mathbf F_{\mu\nu}\Big)\Big]\,,
\end{align}
where tr is a trace over the gauge-group indices, and $\mathbf F$ denotes the (non-Abelian in general) gauge field strengths that enter the chiral anomaly. The term inside the square bracket $\Big[\dots \Big]$ is a total derivative, and expresses the mixed anomaly. 
In the StRVM, in the early universe, gauge fields have not been generated before the end of the RVM inflation. 
Nonetheless, to understand how we can define the KR axion coupling, we need to discuss the coupling of the $b$ field to the chiral anomaly in a gauge sector. 
In this setting, the axion coupling $f_b$ is defined in such a way that 
the axion $b$ would couple to the chiral anomaly (in the gauge sector of the effective action) with standard normalisation,  as $\int d^4x \sqrt{-g} \, \frac{1}{16\pi^2\, f_b}\, \rm tr \Big(\mathbf F^{\mu\nu}\, \mathbf F_{\mu\nu}\Big)$. Given the topological result for the Pontryagin index~\cite{eguchi}  $\frac{1}{16\pi^2} \int d^4x \, \rm tr \Big( \mathbf F^{\mu\nu}\, \mathbf F_{\mu\nu}\Big) \in \mathbb Z$, where $\mathbb Z$ the set of integers, 
 we observe that the effective shift-symmetry-breaking potential, induced by gauge-group instantons, 
is periodic under the shifts $b(x) \to b(x) + 2\pi f_b$. 

Note that the world-sheet-instanton-induced, non-perturbative, periodic modulation terms in \eqref{effpot} do {\it not} generate axion masses.
 In the analysis of \cite{Mavromatos:2022yql}, the energy scales $\Lambda_1, \Lambda_2 $ associated with the 
 periodic modulation terms of the effective axion potential, as well as the 
 parameters $\tilde \xi_1, f_a$, are phenomenological, but with the constraint that:
 \begin{align}\label{constr}
 \Big(\frac{f_b}{f_a} + \frac{\Lambda^4_2}{f_a\, \Lambda^3_0} \Big)^{1/3}\, \Lambda_0  \, < \, \Lambda_1 \ll \Lambda_0 ~.
 \end{align}
 The latter is necessary to ensure that the RVM inflation is driven by the $b(x)$ field, which is the essence of the StRVM.
 The compactification-axion $a(x)$ field, on the other hand, which develops periodic modulations in its potential, that dominate late stages of RVM inflation, is responsible for only prolonging the RVM inflation. 
 
\begin{figure}[ht]
\centering
\includegraphics[width=45mm,height=50mm]{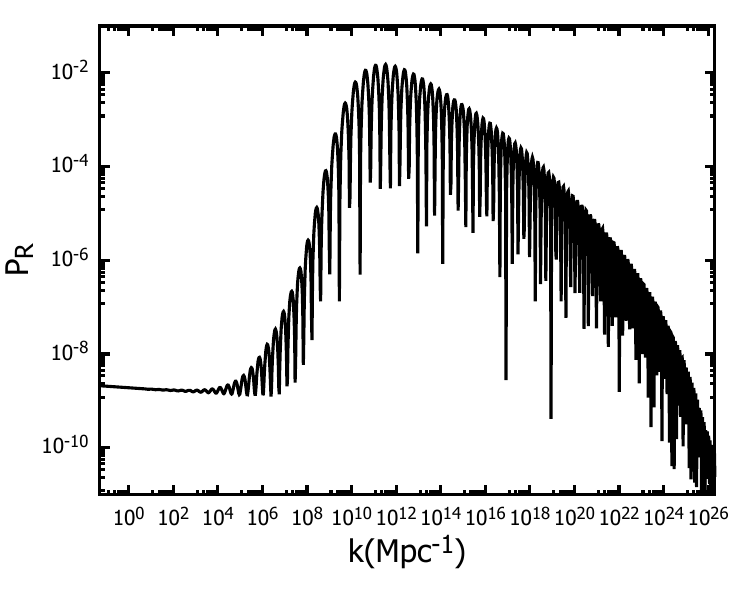}
\includegraphics[width=45mm,height= 55mm]{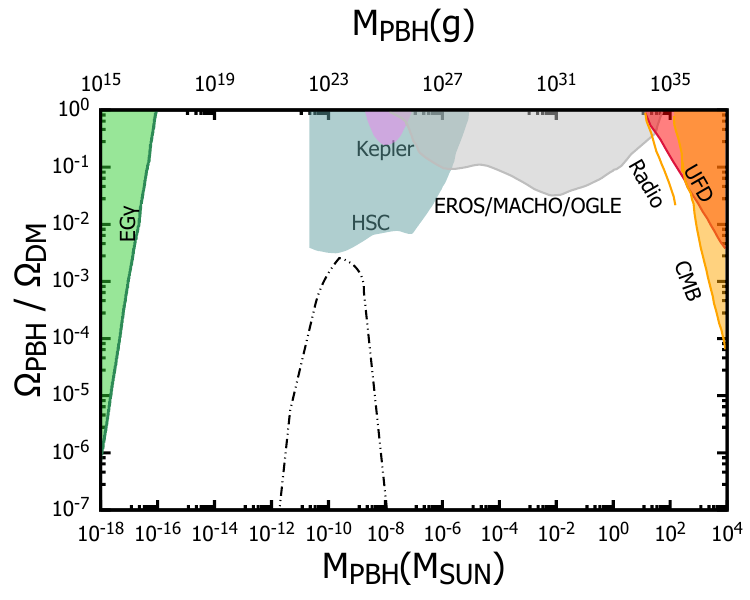} 
\includegraphics[width=55mm]{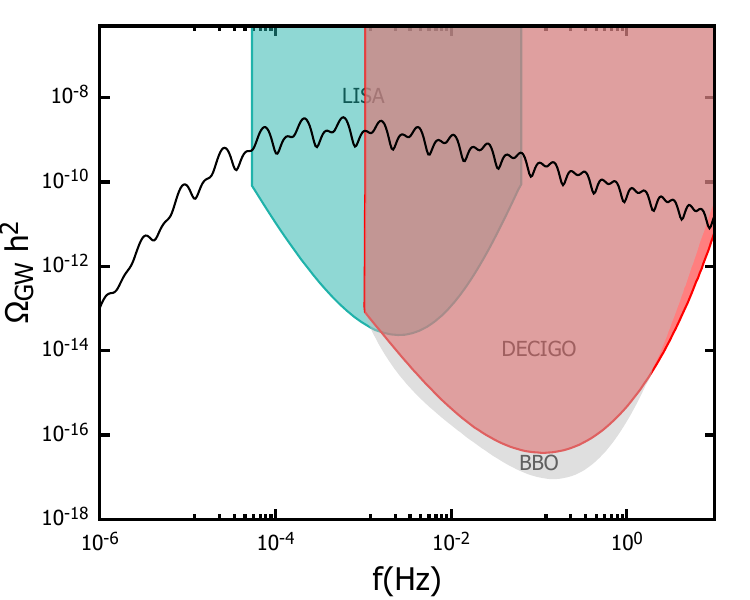} \\
\includegraphics[width=45mm,height= 50mm]{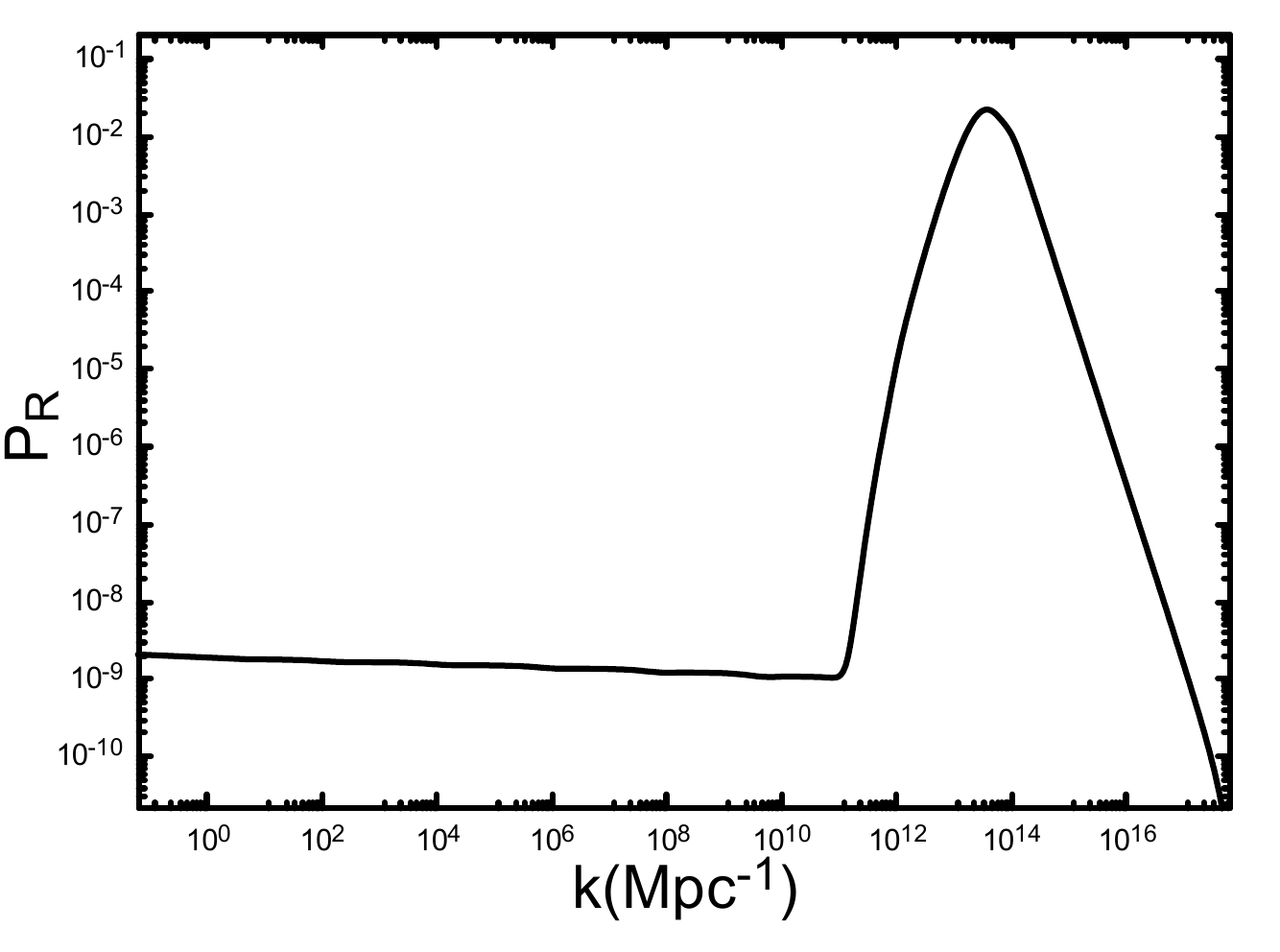}
\includegraphics[width=45mm,height= 55mm]{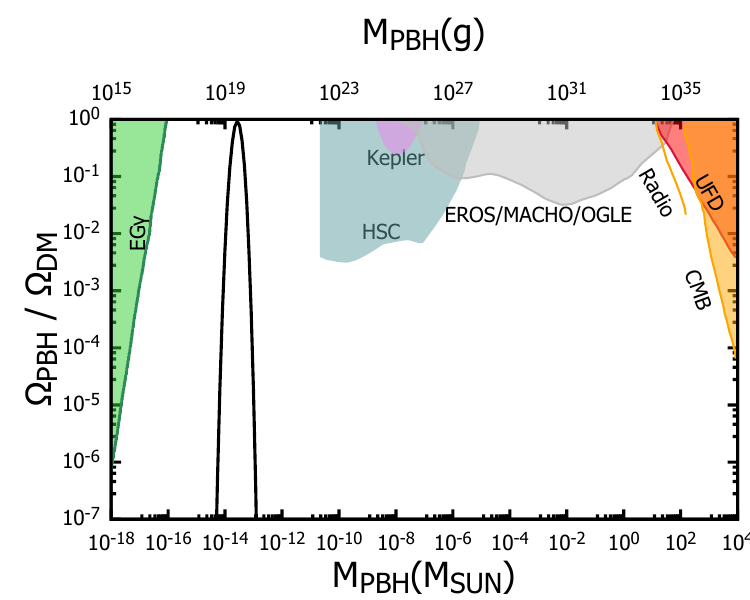}
\includegraphics[width=55mm]{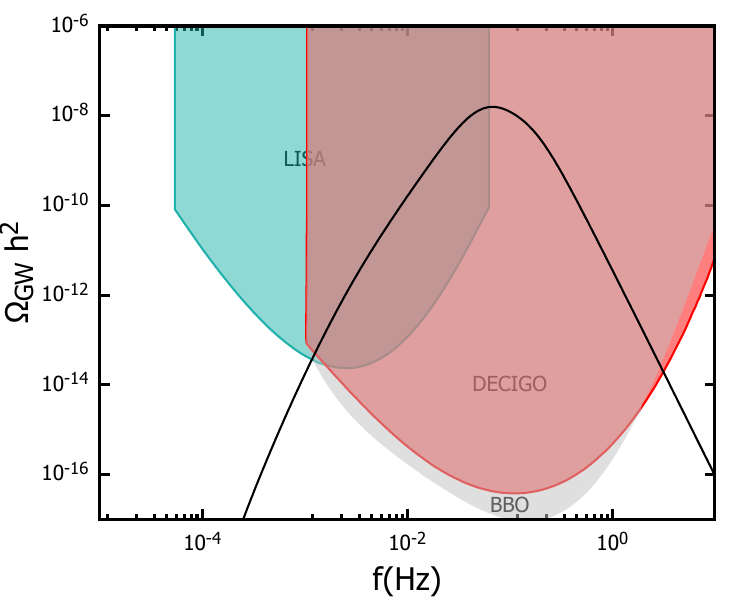}
\caption{The results of the power spectrum ($P_R$), as a function of the energy scale $k({\rm Mpc}^{-1})$, fractional abundance of PBHs of mass $M_{\rm PBH}$
($\Omega_{\rm PBH}/\Omega_{\rm DM}$), and the energy density of induced GWs of frequency $f(Hz)$ ($\Omega_{\rm GW}\, h^2$), for some representative set of parameters, and appropriate initial conditions for inflation: {\it \underline{Upper three figures}}  in the context of the StRVM-inflationary model, with the scale hierarchy \eqref{constr}, as 
given in \cite{Mavromatos:2022yql}, which these figures have been taken from. For concreteness, the fractional abundance of pBH is $f_ {PBH}=0.01$. {\it \underline{Lower three figures}}: the same as above, but for the model of \cite{Zhou:2020kkf}, with the hierarchy of scales \eqref{constr3}. The parameters, which these figures correspond to can be found in \cite{Mavromatos:2022yql}. We also show experimental regions already, or to be, excluded, including those from future interferometers (such as LISA)}.
\label{fig:power}
\end{figure}

Par contrast, in conventional string model linear-axion inflation~\cite{Zhou:2020kkf},  the opposite situation is realised, and the inflation is driven by the compactification axion, whose potential contains the world-sheet-instanton-induced periodic structures, while the KR axion is responsible simply for prolonging - via its linear potential - the duration of inflation, which however is not of RVM type. In this case, the following hierarchy of scales is valid:
 \begin{align}\label{constr3}
 \Lambda_0 \, \ll  \, \Big(\frac{f_b}{f_a} + \frac{\Lambda^4_2}{f_a\, \Lambda^3_0} \Big)^{1/3}\, \Lambda_0  \, < \, \Lambda_1 ~.
 \end{align}
The situation on the effects of the enhanced PBH density on the profiles of the GW in the two cases is given in 
fig.~\ref{fig:power}.  Although both cases \eqref{constr}, \eqref{constr3} lead to enhanced GW power spectra and enhanced production of pBH during inflation, nonetheless the conventional string-monodromy case \eqref{constr3}
is characterised by featureless energy density profiles of GW after inflation, in contrast to the case \eqref{constr}, where the periodic modulations of the compactification-axion-potential at the end of the KR-axion-driven RVM inflation lead to 
a rich structure in the GW profiles during the radiation epoch, experimentally distinguishable  in principle, {\it e.g.} in future interferometers, such as LISA~\cite{Mavromatos:2022yql}.

The above discussion, therefore, points to the fact that stringy axions can play an important r\^ole in early-Universe cosmology, not only by driving RVM-type inflation, through GW condensation of gCS terms, but also by enhancing the density of PBHs, thus increasing the possibility of modified GW profiles during early radiation. Moreover, through the axion-induced increased PBH densities mechanism, described above, 
they also imply potential contributions to Dark Matter (DM), provided the PBHs have masses in the appropriate windows to play the r\^ole of such DM components. In fact, as explained in \cite{Mavromatos:2022yql}, this axion-induced amplification mechanism allows for potentially significant amounts of PBH-DM in the Universe, for some ranges of the parameters of the models.

Axions, however, can also play an important r\^ole in post-RVM-inflationary epochs of the StRVM cosmology, including a direct contribution to DM by the stringy axions upon mass generation, but also to the lepton asymmetry in theories which have massive right-handed neutrinos (RHN) in their spectra. We proceed to discuss such features in the next section.

\section{Post-Inflationary Epochs and quantum-gravity corrections: dark matter of geometric origin and the potential alleviation of the cosmological tensions}\label{sec:tensions}

The presence of CS gravitational anomalous terms during the post inflationary epoch could potentially affect significantly the cosmological evolution, including Big-Bang nucleosynthesis (BBN) data, if such terms survive inflation.
As already mentioned in the previous section, the GW-induced condensates  \eqref{CScond}, do survive the exponential expansion during the RVM inflation, provided a sufficiently large number of GW sources exist in the primordial pre-RVM inflationary universe~\cite{Mavromatos:2022xdo}. However, in the model of \cite{Basilakos:2019acj,Mavromatos:2020kzj}, at the final stages of the RVM inflation, the decay of the metastable vacuum produces among other radiation and matter fields, also chiral fermionic matter fields. The presence of the latter, imply additional terms in the effective Lagrangian \eqref{sea3}, of the form~\cite{Basilakos:2019acj}:
\begin{align}
S_F^{{\rm eff}\,({\rm end\, I})} \ni {\rm kinetic~fermion~terms} + \mathcal M_f\, \int d^4 x\sqrt{-g} \, b \, J^{5\mu}_{\,\,\,\,\,\,;\mu} + {\rm repulsive}~J^{5\mu} J_{\mu}^5~{\rm interactions}
\end{align}
where the semicolon $;$ denotes gravitational covariant derivative, and $\mathcal M_f$ is a coefficient to be determined by the requirement~\cite{Basilakos:2019acj,Mavromatos:2020kzj}
of cancellation of the primordial gCS term in \eqref{sea3}. The current $J^{5\mu} = \sum_i \overline \psi_i \, \gamma^5 \, \gamma^\mu \, \psi_i $ is an axial current summed over all chiral fermion degrees of freedom, including stringy ones.
In the presence of (massless) chiral fermions circulating in quantum loops, the axial current is not conserved at a quantum level. The well-known exact result of the one-loop computation of the chiral anomaly yields terms for the covariant divergence of  $J^{5\mu}$ which are of a mixed anomaly term for curved spacetimes, as the ones pertaining to Cosmology: 
\begin{align}\label{divJ5}
J^{5\mu}_{\,\,\,\,\,\,;\mu} =  c_1 R_{\mu\nu\rho\sigma} \widetilde{R}^{\nu\mu\rho\sigma}  + c_2\, \mathbf F_{\mu\nu} \, \mathbf F^{\mu\nu}\,,
\end{align}
where $c_i, i=1,2$ are numerical coefficients, which are proportional to the number of chiral fermion species, circulating in the loop of the chiral anomaly. According to the requirement of gravitational anomaly cancellation during the post-inflationary eras, postulated in \cite{Basilakos:2019acj,Mavromatos:2020kzj}, the number of chiral fermion species is tuned
in such a way so that the primordial gCS term in \eqref{sea3} is {\it cancelled} by the corresponding term on the right-hand-side of Eq.~\eqref{divJ5}. Only chiral anomalies in the gauge sector survive. During the epoch in which the dominant energy scale is that of QCD, the latter yield, through the instanton effects of the SU(3) colour-group, shift-symmetry-breaking periodic potential for the KR axion of the form~\cite{Basilakos:2020qmu}:
\begin{align}\label{krmass}
V_b = \Lambda_{\rm QCD}^4 \Big(1 - {\rm cos}\big(\frac{b}{f_b}\big)\Big)\,,
\end{align}
where $\lambda_{\rm QCD}$ is the QCD energy scale of order  200 MeV, and $f_b$ the KR axion coupling, defined earlier
\eqref{kraxcoupl}. The potential \eqref{krmass} contains a (non-perturbatively generated) mass term for the KR axion, of order $m_b = \frac{\Lambda_{\rm QCD}^2}{f_b} $, implying that in such scenarios the KR axion can play the r\^ole of a DM component. The order of such a mass depends on the relative magnitude of the string scale $M_s$ to the Planck scale $M_{\rm Pl}$. 
The maximum allowed value of $M_s \sim M_{\rm Pl}$ yields a lower bound on  $m_b \gtrsim 10^{-11}$~eV. For lower values of $M_s$ one gets QCD axion masses within the expected range for non-thermal production using misalignment mechanisms~\cite{Adams:2022pbo}).

\begin{figure}[ht]
\centering
\includegraphics{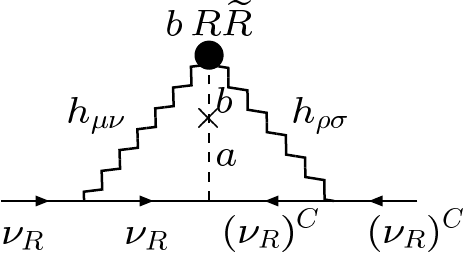}
\caption{Radiative mechanism for Majorana RHN ($\nu_R$) mass generation through shift-symmetry breaking axion-RHN couplings, in the presence of Chern-Simons gravity. The suffix $C$ indicates conjugate neutrino state. $h_{\mu\nu}$ indicate quantum graviton fluctuations about a space-time background, which can be taken to be the FLRW expanding universe spacetime. The dark blob denotes the gCS term coupled with the KR axion. The ``X'' indicates kinetic mixing of the KR axion to the axion from compactification. Picture taken from \cite{pilaftsis}.}.
\label{fig:RHNaxions}
\end{figure}

However, there is an additional mechanism for mass generation of the initially massless stringy axions, in post inflationary epochs of models that contain massive RHN. In principle, at the end of the RVM inflation, massless RHN are also generated, alongside the rest of the chiral fermions. RHN can themselves couple to axions via shift-symmetry breaking non perturbative stringy effects, and acquire mass according to the two-loop anomalous mechanism of \cite{pilaftsis}. According to this scenario, which applies to a  string-inspired model with a KR and another axion field, coming, say, from compactification, two gravitons emanating from a gCS term couple to RHN, together with the KR axion which mixes kinetically with the axion from compactification. The latter then couples directly to the RHN, via chirality changing non-perturbative couplings (see fig.~\ref{fig:RHNaxions}). 
In our scenario, we may assume that such a Majorana RHN mass term is generated at the early phase of radiation. soon after the exit from RVM inflation. The resulting mass may be assumed sufficiently heavy (for details of the mass see discussion in \cite{pilaftsis}). 

Integrating out in the effective theory such massive RHN (which do not respect the shift symmetry) results in an effective shift-symmetry-breaking potential and mass terms for the respective axions. For instance, as discussed in \cite{sarkaraxion}, 
the interaction of the massive RHN with initially massless axions, such as the KR axion, has the generic structure
\begin{align}
\label{interaction}
\bar{N} \  \left( \slashed p-m_{N}-X[\tilde{b} ]\right)  N,
\end{align}
where $X$ can be parametrised as~\cite{Ellis:2020ivx}:
\begin{align}
X\left[ \tilde{b} \right]  =W_{0}\left[ \tilde{b} \right]  +iW_{1}\left[ \tilde{b} \right]  \gamma^{5} +V_{\mu }\left[ \tilde{b} \right]  \gamma^{\mu } +A_{\mu }\left[ \tilde{b} \right]  \gamma^{\mu } \gamma^{5}. 
\end{align}
In our case of interest, the relevant part of $X\left[ \tilde{b} \right]$ is 
\begin{align}
W_{1}\left[ \widetilde{b} \right]  =\frac{2m_{N}}{f_{b}}  \widetilde{b}.
\end{align}
The resultant effective axion potential (up to dimension $6$ operators, which we restrict our attention to, for the purposes of this talk) is:
\begin{align}
\label{effpotb}
V_{\rm eff} [b] = a_{2}\left( W_{1}\left[ \widetilde{b} \right]  \right)^{2}  +a_{4}\left( W_{1}\left[ \widetilde{b} \right]  \right)^{4}  +a_{6}\left( W_{1}\left[ \widetilde{b} \right]  \right)^{6}   + \dots \,,
\end{align}
where 
\begin{align}
\label{acoeffs}
a_{2}=4m^{2}_{N}\left( 1-\frac{1}{2} \ln \frac{m^{2}_{N}}{\mu^{2} } \right) \,, \quad   a_{4}=\frac{5}{6} -\ln \frac{m^{2}_{N}}{\mu^{2} } \quad  {\rm and} \quad 
  a_{6}=-\frac{1}{3m^{2}_{N}}\, .
  \end{align}
  In the above formulae, $\mu$ is a dimensional-regularisation-induced transmutation mass scale 
  in the effective theory. The latter is defined below an UV cutoff scale $\Lambda$, which is determined by the heavy RHN mass scale $m_N$. For the validity of the effective theory 
  we need to have for the KR axion coupling $f_{b} \gtrsim m_N$. We note, for completeness, that, if the heavy neutrinos are used in Leptogenesis, according to the mechanism of \cite{Mavromatos:2012ii,deCesare:2014dga,Bossingham:2017gtm,Bossingham:2018ivs,Mavromatos:2020dkh}, which also applies to the StRVM framework~\cite{Basilakos:2019acj,Mavromatos:2020kzj} in the early radiation era, a plausible 
order of the RHN mass  is $m_N \sim 10^5$~GeV. Of course, much heavier RHN are also allowed, upon appropriate 
choice of the parameters of portal interactions between the RHN and standard model sectors.
In our case, as already mentioned, the large mass $m_N$  plays the r\^ole of an ultraviolet cutoff for the effective theory.  One may do the following matching~\cite{Kaplan:2005es}  at 
 \begin{align}\label{mN=mu}  
  m_N \simeq \mu\,,
  \end{align}
 and run the scale $\mu$ in the regime 
  \begin{align}\label{mNmu} 
  \mu \lesssim m_N \,,
  \end{align}
to lower values until one hits an experimentally measurable physical mass scale.~In our system, such a mass scale would be the axion $b$ mass $m_b$. The latter is generated by the integration of the heavy RHN, which induces an effective KR axion mass given by: 
 \begin{align}\label{axionmass}
  m_b^2 = 32\, \frac{m_N^4}{f_b^2}\, \Big(1 - \frac{1}{2}{\rm ln}[\frac{m_N^2}{\mu^2}]\Big)\,.
  \end{align}
  
      For the string model \eqref{sea3}, we obtain, from \eqref{axionmass} and  \eqref{kraxcoupl}, the following mass for the axion $b$:
      \begin{align}\label{massbvalue}
      m_b \sim 4.35 \, \frac{m_N^2 \, M_{\rm Pl}}{M_s^2}\,.
      \end{align}
 Thus, for the maximum allowed value of $ f_b \sim \, \kappa^{-1} $ (for the case $\alpha^\prime \sim \kappa^2$), upon 
 fixing $\mu \sim m_N \ll f_b$ to order at least $10^5$~GeV, as in the leptogenesis model of \cite{deCesare:2014dga,Bossingham:2017gtm,Bossingham:2018ivs,Mavromatos:2020dkh}), we obtain from \eqref{axionmass} axion masses of minimum order 
$m_b \gtrsim$ 18~eV. This mass is \emph{too large} for the KR axion in this model to play the role of the QCD axion (which is expected to have mass of $\mathcal O(10)~\rm{\mu eV}$ at most~\cite{Adams:2022pbo}).  However, 
this mass is within the range of an axion-like-particle (ALP) \emph{dark matter} component, including those coming from detailed string models after appropriate compactification to (3+1)-dimensions~\cite{Marsh:2015xka,Mehta:2021pwf,Mehta:2020kwu}. Indeed, an acceptable ALP mass range, consistent with astrophysical and cosmological constraints~\cite{Marsh:2015xka,Raffelt,Arvanitaki1,Arvanitaki2,Adams:2022pbo}, is: $10^{-20} \, {\rm eV} \lesssim m_{\rm ALP} \lesssim {\rm eV}$, where the upper bound in this relation pertains to 
ALPs produced by simple misalignment mechanisms, which -unlike  the QCD axions - do not couple directly to gluons~\cite{Adams:2022pbo}. In general, astrophysical constraints allow for much heavier ALP masses.
The lower bound in the above range of ALP masses requires that the axion constitutes the whole of the dark matter. The  models in  \cite{Basilakos:2019acj,Mavromatos:2020kzj} are not restrictive on this last requirement, though, and the axions could be only part of DM. Hence, some of the above constraints are relaxed. We note that, due to \eqref{kraxcoupl}, \eqref{massbvalue}, the existence of an experimentally probed upper bound (of order of eV) on the ALP mass,
allowed by simple misalignment mechanisms,  if applicable to our KR axion in the StRVM case, is quite restrictive for the string scale in such models, allowing, for fixed values of $m_N$, only models with string scales  $ M_{\rm Pl} \ge M_s > 2.1 \, m_N \, \Big(\frac{M_{\rm Pl}}{\rm eV}\Big)^{1/2} \sim  10^{14} \, m_N $.  Hence, for the range of $m_N \gtrsim 10^5$~GeV
of the leptogenesis model of \cite{deCesare:2014dga,Bossingham:2017gtm,Bossingham:2018ivs,Mavromatos:2020dkh}, 
this is excluded, as it would imply $M_s $ larger than $M_{\rm Pl}$. Heavier axion masses are of course allowed in such models, which points to different than misalignment mechanisms for the production of such axions.

Another reason for preference in such large string scales comes from the stability of the vacuum of models with heavy RHN.  Indeed, the ground state (vacuum) of the effective potential \eqref{effpotb} is in general metastable. In \cite{sarkaraxion}, estimates of the life time $\tau$ of the false vacuum have been derived, using the well-known method of bounces~\cite{coleman1,coleman2,coleman3},  with the result that 
\begin{align}\label{lifetimemu=mN}
\frac{\tau}{T_U} &\sim (m_N T_U)^{-4} \, \exp(\frac{8\pi^2}{3|\lambda^{(E)}_{\rm eff}(\eta, \mu=m_N)|}) \, \simeq \, 6.5  \times 10^{-240} \, \Big(\frac{M_{\rm Pl}}{m_N}\Big)^4 \, \exp\Big[\frac{8\pi^2}{51.9} \Big(\frac{f_b}{m_N}\Big)^4\Big]  \nonumber \\
&\simeq 1.6 \times 10^{-185} \, \exp\Big[1.5 \Big(\frac{f_b}{m_N}\Big)^4\Big]\,,
\end{align}
where we took into account that the Age of the Universe $T_U \sim 1.6 \times 10^{60}\, M_{\rm Pl}^{-1}$, and in the last line we have fixed $m_N \sim 10^5$~GeV, using the model of leptogenesis of \cite{deCesare:2014dga,Bossingham:2017gtm,Bossingham:2018ivs,Mavromatos:2020dkh}, for concreteness, which the StRVM is compatible with. To arrive at  the above formula, we have represented the interaction part of the effective potential \eqref{effpotb} in Minkowski spacetime as~\cite{sarkaraxion} 
\begin{align}\label{effpotblambda}
 V^{\rm int}_{\rm eff}[b] \simeq \frac{\lambda_{\rm eff}(b, \mu)}{4}\, b^4 \,, \quad \lambda_{\rm eff}(b, \mu) \equiv  \frac{2^6\, m_N^4}{f_b^4}\, \Big[ \frac{5}{6}- {\rm ln}(\frac{m_N^2}{\mu^2})  - \frac{4}{3\, f_b^2}\, b^2 \Big]\,,
 \end{align}
 and then passed onto Euclidean (E)) formalism, as required in the standard treatment of bounces~\cite{coleman1,coleman2,coleman3}, by reversing the sign of the effective coupling
 $\lambda_{\rm eff} \to  \lambda^{\rm (E)}_{\rm eff} = - \lambda_{\rm eff}$. The quantity $\eta$ appearing in \eqref{lifetimemu=mN} corresponds to the position of 
 the local unstable maxima of the Euclidean effective potential, at  $b= \pm \eta$.
From the result \eqref{lifetimemu=mN} we observe that for $f_b \sim m_N$, the vacuum is highly unstable, while stability is significantly reinforced in models with 
$m_N \ll f_b$, such as those in \cite{deCesare:2014dga,Bossingham:2017gtm,Bossingham:2018ivs,Mavromatos:2020dkh}, in which cases the life time of the vacuum is significantly higher (by several orders of magnitude) than the Universe Age. 

As a last, but not of least importance, topic we now come to discuss the r\^ole of (weak) quantum-gravity corrections in the current cosmological era, where the universe enters again a de-Sitter-like phase. Considering path integration at one loop for graviton fluctuations around a de Sitter cosmological spacetime background, yields in the modern era~\cite{mavrophil,Gomez-Valent:2023hov} a modified gravity effective theory, with logarithmic corrections in the Hubble parameter of the form $H^2 \ln (H)$, which modify the $H^2$ terms coming from the Einstein-Hilbert scalar curvature term. Specifically, as discussed in detail in \cite{Gomez-Valent:2023hov}  (and references therein, where we refer the interested reader for  details), 
the one-loop quantum gravity corrections in the late Universe, about approximately de Sitter spacetimes, may be parametrised by the effective action:
\begin{align}\label{eq:action}
S=-\int d^4x\,\sqrt{-g}\,\left[c_0+ R\, \left(c_1+c_2\ln\left(\frac{R}{R_0}\right)\right)\right]+S_m\,,
\end{align}
where $R_0=12\, H_0^2$ is the de Sitter curvature scalar today, and $S_m$ is a matter/radiation action. The parameter $c_1 = 1/(2\kappa^2) + \delta c_1$, and $\delta c_1$ describes quantum-graviton corrections to the Newton's constant. The sign of $\delta c_1$ may be negative. Unitarity in the gravity sector requires $c_1 > 0$ in our conventions, which is guaranteed due to the weakness of the quantum-gravity corrections. The important point in the analysis of \cite{Gomez-Valent:2023hov} is that the logarithmic term can contribute towards the alleviation of both types of cosmological tensions in the current data, that is $H_0$ and growth-of-structure tensions~\cite{Abdalla:2022yfr}.
\begin{figure}[ht]
\center
\includegraphics[width=35pc]{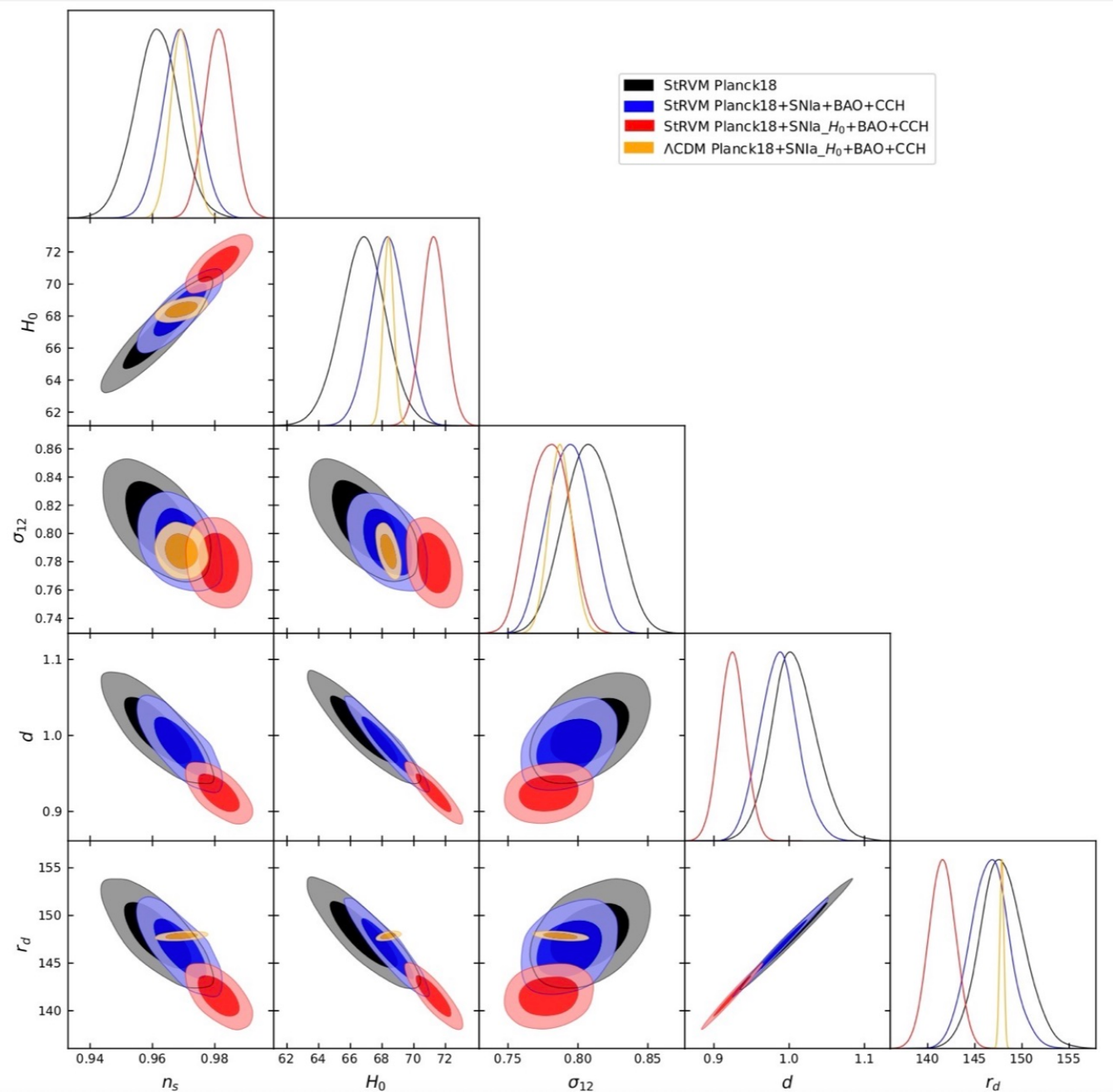}
\caption{The alleviation of the $H_0$ and growth tensions in StRVM, upon the inclusion of graviton quantum fluctuations at one-loop order in modern eras (see \eqref{eq:action}). Here we quantify the amplitude of the matter power spectrum at linear scales by means of the root-mean-square mass fluctuations at length scales of $12$ Mpc, that is, the parameter $\sigma_{12}=\sigma(r=12\,{\rm Mpc})$ . The reasons why $\sigma_{12}$ is a more appropriate parameter than $\sigma_8=\sigma(r=8h^{-1} {\rm Mpc})$ for the phenomenology of the StRVM are explained in \cite{Gomez-Valent:2023hov}, which this figure is taken from. The $\Lambda$CDM predictions are depicted in orange colour.}
\label{fig:strvmH0}
\end{figure}
An  analysis of this class of StRVM, fitting the model to the currently available cosmological data, that is CMB, 
supernovae, Baryon Acoustic Oscillations and cosmic chronometers, has been given in ref.~\cite{Gomez-Valent:2023hov}, where we refer the interested reader for details. Both types of cosmological tensions can be alleviated simultaneously, by selecting appropriate values for the relevant parameters, which notably are consistent with those obtained through BBN analysis in \cite{bbnrvm}.  The presence of the logarithmic curvature corrections in \eqref{eq:action} are crucial to this effect.

We also remark for completion that, in the context of the StRVM, one can make the specific assumption~\cite{Gomez-Valent:2023hov} that the supergravity contributions, corresponding to a primordial supersymmetry breaking scale $\sqrt{|f|}$, which are responsible for a {\it negative} value of the bare cosmological constant  $-f^2 < 0$ (as required by supersymmetry~\cite{houston}), dominate all the quantum-graviton-induced logarithmic corrections to the energy density from the primordial to the current era. In that case, one obtains~\cite{mavrophil,Mavromatos:2020kzj}: 
\begin{align}\label{c12}
c_1 - c_2\, {\rm ln}(\kappa^2\, H_0^2) &=
\frac{1}{2\kappa^2} \Big[1 + \frac{1}{2}\, \kappa^4 \, f^2 \,\Big(0.083 - 0.049 \, {\rm ln}(3\kappa^4\, f^2)\Big)\Big]\, , \nonumber \\
c_2 &= - 0.0045\, \kappa^2 \, f^2 < 0\,.
\end{align}
We then constrain the constants $c_1,c_2$ from data~\cite{Gomez-Valent:2023hov}, in order to alleviate simultaneously the $H_0$ and growth tensions. The situation is summarised
in figure \ref{fig:strvmH0}, for the values of the dimensionless parameters 
\begin{align}\label{param}
|\frac{c_2}{c_1 + c_2}| =
9 \times 10^{-3} \, \kappa^4 \, f^2 \lesssim \mathcal O(10^{-7})\,, \quad 2\kappa^2\, (c_1 + c_2) = 0.924 \pm 0.017\,.
\end{align}
This yields the following estimate on the magnitude of $\sqrt{|f|}$, 
\begin{align}\label{valf}
\sqrt{|f|} \kappa \sim 10^{-5/4} \lesssim 1\,, 
\end{align}
which in turn implies a sub-planckian (but close to the reduced Planck) scale for the primordial dynamical supergravity breaking. This is consistent with the transplanckian censorship conjecture (that is, broadly speaking, the assumption that no physical scale 
exceeds the Planck scale). 

From \eqref{c12}, one also observes that such values are consistent with the perturbative deviations of $c_1$ from the (3+1)-dimensional gravitational constant $1/(2\kappa^2)$, despite the tiny value of the quantity $\kappa^2 H_0^2 = \mathcal O(10^{-122}) \ll 1$. We also note that the quantum-graviton-induced corrections during the RVM inflationary eras, which assume the form $H^4\ln H$, do not affect the order of magnitude estimates of the GW-induced condensate of the gCS terms that lead to RVM inflation~\cite{,Mavromatos:2020kzj}, and, hence, the mechanism for inflation discussed in \cite{Basilakos:2019acj,Mavromatos:2020kzj} and reviewed here.

\section{Conclusions and outlook}\label{sec:concl}

In this talk, we have discussed the various r\^oles axions that characterise string-inspired cosmological models can play in various epochs of the Universe. Specifically, we have seen how axions from the massless gravitational multiplet of strings, can lead to RVM-like inflation, through their interactions with gravitational CS terms, upon condensation of the latter due to primordial chiral GW. Moreover, the so-induced, linear in cosmic time, axion backgrounds survive the exit from the inflationary era and remain intact into the early radiation cosmological epoch. In theories with massive right-handed neutrinos, then, such axion backgrounds cause leptogenesis, which through B-L conserving sphaleron processes in the standard model sector of the theories, can lead to baryogenesis, and thus matter-antimatter asymmetry in the Universe.

In addition to the above r\^ole, axions in strings can also play that of the dominant components of dark matter.
The KR axion, which during inflation is massless, can acquire a non trivial mass during early radiation era, in an epoch where QCD instanton effects are dominant. The latter can generate periodic structures in the KR axion potential, through its interactions with the chiral anomaly of the SU(3) colour group, which are generated after the RVM inflation in the StRVM~\cite{Basilakos:2020qmu}. There is, however, an alternative mechanism for generating KR axion masses, which proceeds via the effective action after integration of heavy Majorana RHN that may characterise the matter quantum-field theory models after the RVM inflation. The presence of the RHN mass breaks the axion shift-symmetry explicitly, with the result that the effective action obtained after the heavy RHN path-integration contains axion mass terms. The vacuum in such effective field theories is metastable. However, for relatively large string scales $M_s$, determined by the magnitude of the RHN mass, the life time of this vacuum is several orders of magnitude longer than the observable Universe age. 
The UV cutoff scale for the validity of the effective action is the mass of the heavy RHN. The latter could also  acquire their masses radiatively, as already mentioned in the previous section, via their non-perturbatively generated shift-symmetry breaking coupling with the KR axions, as well as their gravitational coupling to the gCS term, according to the mechanism proposed in \cite{pilaftsis}. 

There are several aspects of the approach that can still be checked, either for theoretical consistency or in observational cosmic searches. One particularly interesting aspect is the r\^ole of the logarithmic in the Hubble parameter  H corrections on primordial eras, after the exit from inflation, where PBHs are produced. As recently argued in \cite{papanik}, one obtains severe restrictions on the values of the coefficients of these logarithmic terms, in order to avoid overproduction of such PBHs. The required upper bound of these coefficients lie several orders of magnitude below the corresponding values 
required for the the alleviation of the cosmological tensions in the modern era, as discussed at the end of the previous section \ref{sec:tensions} ({\it cf.} fig.~\ref{fig:strvmH0}). The origin of such restrictions is associated with the presence of the so-called scalaron mass, that is the mass of the extra polarisation mode that exist in modified gravity $f(R)$ theories, such as the RVM with quantum-graviton-induced logarithmic curvature corrections. On the assumption that such a mode is physical, one should impose the restriction that, if it exists, its mass should be less than the Planck-mass scale, in order for the transplanckian censorship hypothesis to be respected. This leads to the aforementioned constraints. The interpretation of such a rather unexpected result, is that: \textbf{(i)}  either the assumption that the scalaron mode is a physical mode is false, and in fact in this respect we mention that the General Relativity limit corresponds to an infinite mass (non-propagating) scalaron, in which case there are no restrictions on the logarithmic terms from PBH-dominant era, or \textbf{(ii)} while the scalaron mass is physical, nonetheless the form of the quantum-gravity corrections in the PBH dominant era, is not logarithmic, but rather suppressed, due to the fact that such epochs constitute a strong-gravity regime, far from a de-Sitter era, for which the logarithmic corrections have been derived analytically, in a weak (one-loop)-quantum-gravity setting~\cite{houston,houston2,Gomez-Valent:2023hov}. One should derive the quantum-graviton effects for the PBH era, if possible, and then compare to data. Moreover, given that the PBH overproduction can occur in early matter dominated eras of the Unvierse, in such regimes one should also examine the logarithmic terms in H that arise from quantum-matter corrections, along the lines of \cite{rvmqft1,rvmqft2,rvmqft3,rvmqft4}. 

Another issue worthy of further study within the StRVM framework is the better understanding of its late cosmological eras, especially the origin of the (observed) dark energy dominant component in the energy budget of the current-era Universe. 
Given that the StRVM is embeddable in concrete string-theory models, which are UV complete theories, including quantum gravity, it should be compatible with the swampland conjectures~\cite{swamp1,swamp2,swamp3,swamp4}. Hence, any term that resembles a cosmological constant, which according to the observations, seem to characterise the current era, 
should be metastable, characterised by a finite lifetime, and thus disappearing asymptotically in time. This would allow compatibility of the approach with both, the swampland criteria of non-perturbative string theory, and perturbative strings, given that  such metastable dark energy terms would be compatible with the existence of proper asymptotically well-defined perturbative scattering matrices~\cite{scat1,scat2}. 

We mention for completeness at this point that metastable dark energy in the current epoch also characterises~\cite{emnnc} non-critical (Liouville) string cosmologies~\cite{emnnc2}, which are thus compatible with the existence of a well-defined scattering matrix. It is also explicitly known that Liouville-string cosmologies in the presence of brane space-time defects~\cite{emnnc3} can be characterised by a de-Sitter type equation of state despite a mild time dependence on the effective cosmological energy density $\Lambda (t)$ of their vacua in the current era. This is in similar spirit to the situation that characterises the RVM approach to cosmology. The time dependence of the vacuum energy in such brane/string models arises as a result of the recoil of the brane defects during their scatterings with ``string matter''. This induces a non-criticality of the underlying string theory~\cite{emnnc3}. Although in the way initially formulated~\cite{Basilakos:2019acj,Mavromatos:2020kzj}, there is no non-trivial dilaton configuration in the StRVM, unlike the non-critical string  cosmologies~\cite{emnnc2}, and thus one is tempted to conclude that there is no obvious connection between the two approaches, nonetheless one might dare of speculating that there might be a hidden formal relation with non-critical strings worthy of  further exploration, in view of the StRVM metastable-condensate contributions to the dark energy. The above issues, among others, are actively pursued at the moment, and we hope that we shall be able to report some results soon. 

\section*{Acknowledgements} NEM would like to thank the organisers for their kind invitation to give this plenary talk.
This work is supported in part by the UK Science and Technology Facilities research Council (STFC) under the research grants ST/T000759/1 and ST/X000753/1. The work of P.D. is supported by a graduate scholarship from the National Technical University of Athens.
NEM also acknowledges participation in the COST Association Actions CA18108 ``{\it Quantum Gravity Phenomenology in the Multimessenger Approach (QG-MM)}'' and CA21136 “Addressing observational tensions in cosmology with systematics and fundamental physics (CosmoVerse)”.

\bibliographystyle{JHEP}      
\small

\bibliography{CorfuNEM23}

%

\end{document}